\documentclass[aps,pra,twocolumn,showpacs,letterpaper,tighten,float,superscriptaddress,longbibliography,footinbib]{revtex4-1}
\usepackage{amsmath,amssymb,amsthm,amscd,latexsym,epsfig,bm}

\usepackage[colorlinks,bookmarks=false,citecolor=blue,linkcolor=red,urlcolor=blue]{hyperref}
\usepackage{verbatim}

\usepackage{bbold}
\usepackage{booktabs}
\usepackage{longtable}
\usepackage{amstext}
\usepackage{array}
\newcolumntype{L}{>{$}l<{$}} 
\newcolumntype{C}{>{$}c<{$}} 

\def\beq{\begin{equation}}
\def\eeq{\end{equation}}
\def\be{\begin{equation}}
\def\ee{\end{equation}}

\newcommand{\zz}{\mathbb{Z}_2}
\newcommand{\z}{\mathbb{Z}}

\def\f2{{\mathbb F}_2}

\theoremstyle{plain}
\theoremstyle{plain}

\providecommand{\theoremname}{Theorem}
\providecommand{\theoremtextname}{Theorem}

\theoremstyle{plain}
\providecommand{\propositionname}{Proposition}

\begin{document}

\title{4D beyond-cohomology topological phase protected by $C_2$ symmetry and its boundary theories}
\author{Sheng-Jie Huang}
\affiliation{Condensed Matter Theory Center and Joint Quantum Institute, Department of Physics, University of Maryland, College Park, Maryland 20742-4111, USA}
\date{\today}

\begin{abstract}
We study bosonic symmetry protected topological (SPT) phases with $C_{2}$ rotational symmetry in four spatial dimensions which is not captured by the group cohomology classification. By using the topological crystal approach, we show that the topological crystalline state of this SPT phase is given by placing an $E_{8}$ state on the two dimensional rotational invariant plane, which provides a simple physical picture of this phase. Based on this understanding, we show that a variant of QED4 with charge-1 and charge-3 Dirac fermions is a field theoretical description of the three dimensional boundary. We also discuss the connection to symmetric gapped boundary with topological order and its anomaly signature. 
\end{abstract}

\maketitle

\section{Introduction}
Symmetry protected topological phases are gapped phases of matter with a unique ground state which can be adiabatically connected to a trivial product state if the symmetry is broken explicitly\cite{Schnyder2008, Kitaev2009, ryu2010, gu2009, pollmann2010, fidkowski2011, turner2011, chen2011_1dspt, chen2011_1dcomplete, cirac2011, chen2013cohomology, levin2012}. For bosonic systems with onsite symmetry, a large class of SPT phases are classified by group cohomology\cite{chen2013cohomology}. Although group cohomology captures some of the mathematical structure of SPT phases correctly, it has been recognized that the structure of the classification of SPT phases should be described by generalized cohomology theories\cite{Xiong2018,Gaiotto2019}. In particular, there are many ``beyond-cohomology" SPT phases which are not captured by group cohomology. These beyond-cohomology SPT phases are not as well-understood as in-cohomology SPT phases. 

Besides the development of onsite SPT phases, there has been a great progress of SPT phases protected by crystalline symmetry\cite{Fu2011,Hsieh2012,ando2015,Chiu2013,Chiu2016,Shiozaki2014,Isobe2015,Fang2015,Shiozaki2015,Yoshida2015,Wang2016,Fang2017a,Po2017,Bradlyn2017,kruthoff2017,song17topological,jiang17anyon,thorngren18gauging,huang17building,Huang2018surface,Song2018E8,Cheng2018,Gu2019,Shiozaki2018AH,Dominic2019defect,Song2018realspace,Song2019TC}. Crystalline SPT phases are known to have a simple physical picture given by the ``topological crystals", which are a special class of states formed by real-space crystalline patterns of lower-dimensional topological states\cite{song17topological,huang17building,Song2018E8,Shiozaki2018AH,Dominic2019defect,Song2018realspace,Song2019TC}. Based on the topological crystal approach, most crystalline SPT phases are easier to understand compared to SPT phases with onsite symmetry. Moreover, it has also been found that the classification of cSPT phases with a spatial symmetry group $G$ is the same as the classification of SPT phases with onsite symmetry G, which is known as the Crystalline Equivalence Principle\cite{thorngren18gauging}. Sometimes, it's helpful to think about the crystalline counterpart of an onsite SPT phases since the former is usually easier to understand. 

3D beyond-cohomology SPT phases with crystalline symmetry have been classified systematically by using the topological crystal approach\cite{Song2018E8}. Much less is known about the 4D beyond-cohomology SPT phases. For on-site unitary symmetry, one of the simplest beyond-cohomology SPT phase is given in the case of 4D bosonic system protected by onsite $Z_{2}$ symmetry. The response of this phase has been studied at the field theory level\cite{Kapustin2014,Wen2015,Freed2016}. Recently, Ref.~\cite{Fidkowski2019} constructs an exactly solvable model for this phase and studies its quantized response at the Hamiltonian lattice level. The bulk physical picture of this phase is also revealed by their construction: decorating $Z_{2}$ domain walls with 3D Walker-Wang models based on the so-called ``3-fermion" topological order, which is a variant of toric-code topological order where all the anyonic excitations are fermionic. 

In this paper, we are going to study the crystalline counterpart of the 4D beyond-cohomology SPT phase where the onsite $Z_{2}$ symmetry is replaced by $C_{2}$ rotation. We are going to see that the beyond-cohomology SPT phase with $C_{2}$ rotation is much easier to understand by using the topological crystal approach. We then move on to study its 3D boundary theories. We found that one possible boundary field theory is given by a variant of QED4 with charge-1 and charge-3 Dirac fermions. This proposal is supported by applying dimensional reduction argument on the boundary. We also consider how to obtain the anomalous boundary topological order from the field theory we obtained. The anomalous boundary topological order is shown to be a 3D $Z_{2}$ gauge theory with a fermionic gauge charge. This is consistent with the finding in Ref.~\cite{Fidkowski2019} for the case of onsite $Z_{2}$ symmetry. We further show that one of the anomaly signature of the 3D gauge theory is shown in the core of the loop excitation, in which there are gapless modes that are adiabatically connected to the edge modes of the $E_{8}$ state. Another anomaly signature is that the $C_{2}$ defect loop carries gapless modes that are equivalent to $c=4$ $SO(8)_{1}$ chiral CFT. 

\section{Classification}
In this section, we classify bosonic $C_{2}$ SPT phases by using the topological crystal approach. Let $(x,y,z,w)$ be the coordinate of the 4D space. We consider a $C_{2}$ rotational symmetry which acts on the spatial coordinates by $C_{2}: (x,y,z,w) \rightarrow (x,-y,-z,w)$. Since there is no symmetry away from the rotational invariant plane $(x,0,0,w)$, the 4D bulk can be extensively trivialized to a product state except on the rotational invariant plane. On this plane, the $C_{2}$ rotational symmetry becomes a on-site $\zz$ symmetry. This 2D plane could support two possible short range entangled states. The first possibility is the Ising SPT state\cite{levin2012}. This is the single non- trivial SPT phase protected by an onsite $\zz$ symmetry. The other possibility is the $E_{8}$ state, which has a $\z$ classification in 2D. We can take the $\zz$ symmetry to act trivially on the $E_{8}$ root state. From these root states, we obtain a $\zz \times \z$ classification of 2d phases on the rotational invariant plane.

\begin{figure}
\includegraphics[width=1\columnwidth]{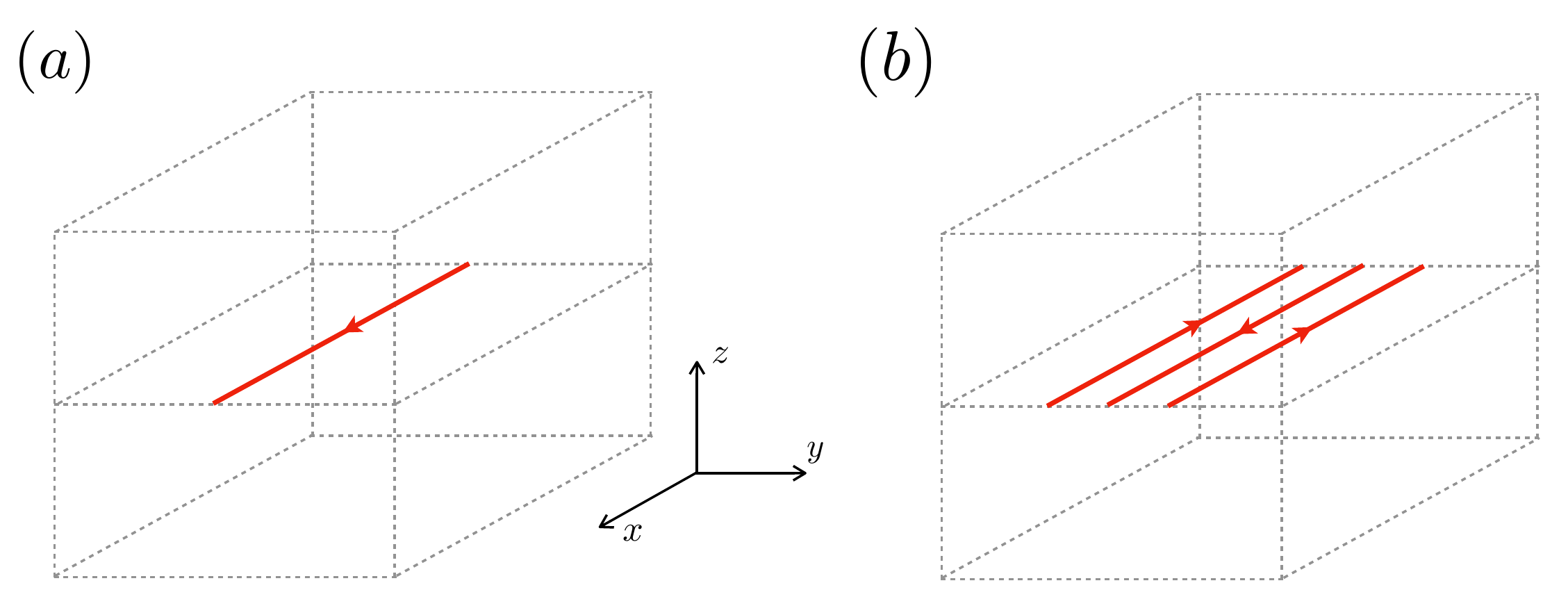}
\caption{(a) The projection of the $E_{8}$ root state to the 3D boundary. The red line depicts the edge modes of the $E_{8}$ state and the arrow shows the chirality. The $E_{8}$ edge modes are pined on the rotational axis due to $C_{2}$ rotational symmetry. (b) Adjoined $E_{8}$ layers related by $C_{2}$ symmetry changes the chirality of the $E_{8}$ state on the rational invariant axis.
}
\label{fig:E8}
\end{figure}

To obtain the classification of $C_{2}$ SPT phases in 4D, we need to consider block-equivalence relations. Let $|\psi_{0} \rangle$ denotes a state on the rotational invariant plane, this state is equivalent to a state with adjoined layers related by $C_{2}$ symmetry. More precisely, we have this equivalence relation:
\begin{equation}
|\psi_{0} \rangle \sim |L\rangle \otimes |\psi_{0} \rangle \otimes |R\rangle,
\label{eqn:adjoin}
\end{equation}
where $C_{2}$ acts by
\begin{equation}
U_{C_{2}} | L \rangle = |R \rangle, U_{C_{2}} | R \rangle = |L \rangle.
\end{equation}
Now suppose there are $n_{E_{8}}$ copies of $E_{8}$ states on the rotational invariant plane. Since $ |L \rangle$ and $ |R \rangle$ can be $E_{8}$ states of the same chirality, the adjoining layers can change the $E_{8}$ index of $|\psi_{0}\rangle$ by $\pm 2$. Therefore, we found that the $E_{8}$ index $n_{E_{8}}$ should only be well-defined modulo $2$. This result suggests that a state with $n_{E_{8}}=2$ could either be the trivial state, or the Ising SPT state. We show this state is trivial in Sec.~\ref{sec:sto}, by analyzing its boundary topological order. Therefore, we obtain a $\zz \times \zz$ classification of bosonic $C_{2}$ SPT phases in 4D. This is consistent with the TQFT classification of 4D SPT phases protected by the on-site $\zz$ symmetry\cite{Fidkowski2019} as one would expected from the crystalline equivalence principal\cite{thorngren18gauging}. 

\section{Boundary field theory}
\label{sec:sft}
Here we discuss a boundary field theory of the $E_{8}$ root state. Our argument starts by considering the 4D $E_{8}$ TCI with $U(1) \times C_{2}$ symmetry, which is a strongly interacting electronic SPT phase. In the dimensional reduction picture, the 4D $E_{8}$ TCI is described by placing a neutral bosonic $E_{8}$ state on the rotational invariant plane, together with a trivial electronic insulator. By gauging the $U(1)$ symmetry, we eliminate the trivial fermionic sector and obtain a bosonic state with an $E_{8}$ state on the rotational invariant plane. We are going to first argue that one possible boundary field theory of the $E_{8}$ TCI is a $(3+1)$d Dirac theory with both charge-1 and charge-3 Dirac fermions. Once we obtain this boundary field theory, we gauge the $U(1)$ symmetry to obtain a theory for the bosonic $E_{8}$ root state. This theory is a variant of $N_{f}=2$ flavors of QED4 with charge-1 and charge- 3 Dirac fermions. 

To obtain the boundary field theory, we are going to consider an alternative description of the $E_{8}$ TCI. Starting from a $E_{8}$ state with $c=-8$ on the rotational invariant plane, we also bring in a $\nu=8$ integer quantum hall (IQH) state on the rotational invariant plane. The resulting state is characterized by $c = 0$ and hall conductance $\nu = 8$. To see that this procedure leaves the system in the same phase, we need to show that putting $\nu = 8$ IQH state alone on the rotational invariant plane is in a trivial phase. Let's begin by consider two $\nu = 1$ IQH state on the rotational invariant plane formed by fermions $c_{1}$ and $c_{2}$. The $C_{2}$ symmetry acts trivially. Then we consider adjoined layers with $|L\rangle$ and $|R\rangle$ being $\nu=-1$ IQH state with fermions $d_{L}$ and $d_{R}$. $C_{2}$ symmetry acts on fermions by exchanging $d_{L}$ and $d_{R}$. We take linear combinations $d_{\pm} = d_{R} \pm d_{L}$ with eigenvalue $\pm1$ under $C_{2}$. We can gap the two IQH states with $d_{+}$ and $c_{2}$ fermions while preserving $C_{2}$ symmetry since $d_{+}$ and $c_{2}$ fermions both have positive eigenvalues under $C_{2}$ and have opposite chirality. This leaves a non-chiral state, where the $c_{1}$ and $d_{-}$ fermions have opposite eigenvalues under $C_{2}$ and form IQH states of opposite chirality. This state is precisely the SPT state with $U(1) \times \zz$ symmetry considered in \cite{Isobe2015}. It was shown that the classification is $\z_{4}$ in the presence of interactions. We thus found that $\nu=8$ IQH state on the rotational invariant plane is equivalent to 4 copies of $U(1) \times \zz$ SPT state, which is in the trivial phase.

We see that the $E_{8}$ TCI can be described by placing a state with $c = 0$ and hall conductance $\nu = 8$. This is equivalent to placing a bosonic integer quantum Hall (BIQH) state on the rotational invariant plane, built from charge- 2 Cooper pairs together with a trivial electronic insulator. To proceed, we use the ``cluston'' construction of the 2D BIQH state\cite{Chong2015cluston}. The idea is to consider binding three electrons into cluston bound states, and then putting the clustons into a Chern band. The Cooper pair BIQH state can be obtained by combining a $\nu=1$ IQH state of clustons with a $\nu = -1$ IQH state of electrons. The resulting state has the desired boundary signature: quantum hall conductance $\nu= 8$ and central charge $c = 0$. This picture suggests the following $(3+1)$d boundary field theory 
\begin{equation}
\mathcal{L} = -i \bar{\psi} \gamma^{\mu} \partial_{\mu} \psi -i \bar{\psi}_{c} \gamma^{\mu} \partial_{\mu} \psi_{c},
\label{eqn:cluston}
\end{equation}
where $\psi$ is a charge-1 fermion and $\psi_{c}$ is a charge-3 cluston. We use the following convention for the gamma matrices:
\begin{eqnarray}
\gamma^{0} &=&
\begin{pmatrix}
\mathbb{1} & 0  \\
0 & -\mathbb{1}  \\ 
\end{pmatrix}
= \tau^{3},
\\
\gamma^{i} &=& 
\begin{pmatrix}
0 & \sigma^{i}  \\
-\sigma^{i} & 0  \\ 
\end{pmatrix}
=i\sigma^{i} \tau^{2},
\end{eqnarray}
and 
\begin{equation}
\gamma^{5} =
\begin{pmatrix}
0 & \mathbb{1}  \\
\mathbb{1} & 0  \\ 
\end{pmatrix}
= \tau^{1}.
\end{equation}
The $(3+1)$d field theory Eq.~\ref{eqn:cluston} is a generalization of the $(2+1)$d field theories discussed in Ref.~\cite{Chong2015cluston,Seiberg2016bdyTI,Meng2016sulfdual}. We are going to argue that Eq.~\ref{eqn:cluston} is the boundary field theory of the 4D $E_{8}$ TCI with the following unconventional $\tilde{C_{2}}$ symmetry:
\begin{eqnarray}
\tilde{C_{2}} : \psi(\bold{r}) &\rightarrow& -i \sigma^{1}\tau^{1} \psi(R\bold{r}),
\label{eqn:3dc2-1}
\\
\tilde{C_{2}} : \psi_{c}(\bold{r}) &\rightarrow&  i\sigma^{1}\tau^{1} \psi_{c}(R\bold{r}),
\label{eqn:3dc2-2}
\end{eqnarray}
where $R\bold{r} = (x,-y,-z)$. This unconventional $\tilde{C_{2}}$ symmetry is a combination of the conventional $C_{2}$ symmetry in the Dirac theory, the charge $U(1)$, and the axial $U(1)_{A}$ symmetry. More precisely, let $u_{C} = e^{i\pi Q/2 }$ and $u_{A}=e^{i\pi \gamma_{5}/2} $ be the generators of the $\z_{4}$ subgroup of the charge $U(1)$ and axial $U(1)_{A}$ symmetry respectively, then $\tilde{C_{2}}= C_{2} u_{C}u_{A}$. It's straightforward to see that Eq.~\ref{eqn:cluston} can't be gapped out by adding spatially uniform mass terms. Note that it's important to include the axial $U(1)_{A}$ symmetry, otherwise Eq.~\ref{eqn:cluston} can be gapped out by a uniform mass term.

Our argument is based on a two-step dimensional reduction procedure. We first add the following mass term
\begin{equation}
\mathcal{L}_{m} = m(z) \bar{\psi} \psi+ m(z) \bar{\psi}_{c} \psi_{c},
\end{equation}
where $m(z)$ is real and satisfies $m(z) = -m(-z)$, and $m(z) \rightarrow m_{0} > 0$ as $z \rightarrow +\infty$. This term preserves all the symmetries. By solving the fermion zero modes on the mass domain wall,  we obtain the gapless fermions on the domain wall that are described by the following $(2+1)$d Dirac theory:
\begin{equation}
\mathcal{L_{D}} = -i \bar{\chi} \tilde{\gamma}^{\mu} \partial_{\mu} \chi -i \bar{\chi}_{c} \tilde{\gamma}^{\mu} \partial_{\mu} \chi_{c},
\label{eqn:cluston_2d}
\end{equation}
with $\tilde{C_{2}}$ acts by
\begin{eqnarray}
\tilde{C_{2}} : \chi(x,y) &\rightarrow& i \tilde{\gamma}^{2} \chi(x,-y),
\label{eqn:2dc2-1}
\\
\tilde{C_{2}} : \chi_{c}(x,y) &\rightarrow& -i \tilde{\gamma}^{2} \chi_{c}(x,-y),
\label{eqn:2dc2-2}
\end{eqnarray}
where $\tilde{\gamma}^{2} = i \sigma^{1}$.

Next, we add the following mass term to Eq.~\ref{eqn:cluston_2d}:
\begin{equation}
\mathcal{L}_{\mathcal{D}m} = -m(y) \bar{\chi} \chi+ m(y) \bar{\chi}_{c} \chi_{c},
\end{equation}
where where $m(y)$ is a real function with $m(y) = -m(-y)$, and $m(y) \rightarrow m_{0} >0$ as $y \rightarrow +\infty$. This results in a pair of counter-propagating chiral modes on the rotational axis, described by the Hamiltonian density
\begin{equation}
\mathcal{H}_{x} = iv_{F} \phi^{\dagger} \partial_{x} \phi - i v_{F} \phi_{c}^{\dagger} \partial_{x} \phi_{c},
\label{eqn:1dchiral}
\end{equation}
where $\phi$ carries charge-1 and $\phi_{c}$ carries charge-3. $\tilde{C_{2}}$ acts trivially on $\phi$ and $\phi_{c}$. This is exactly the edge theory of a Cooper pair BIQH state discussed in Ref.~\cite{Chong2015cluston}. It follows that the state on the rotational invariant plane has $\nu=8$ and $c=0$. Therefore, Eq.~\ref{eqn:cluston} is a boundary field theory for the $E_{8}$ TCI.

We can now gauge the $U(1)$ symmetry to obtain a boundary field theory for the bosonic $E_{8}$ root state. The gauged Lagrangian has the following form
\begin{equation}
\mathcal{L}_{g} = -i \bar{\psi} \gamma^{\mu} (\partial_{\mu}+ia_{\mu}) \psi -i \bar{\psi}_{c} \gamma^{\mu} (\partial_{\mu} + i 3a_{\mu}) \psi_{c} + ... .
\label{eqn:eqn:cluston_gauge}
\end{equation}

\section{Boundary topological order}
\label{sec:sto}
Here we discuss the boundary topological orders of the $E_{8}$ root state. One direct route to enter the topologically ordered state is by Higgsing the $U(1)$ gauge symmetry down to $Z_{2}$. This can be achieved by adding appropriate pairing term to Eq.~\ref{eqn:eqn:cluston_gauge} so that the fermions are in the superconducting state while preserving the $C_{2}$ symmetry. The resulting 3D topological orders we obtain is a $Z_{2}$ gauge theory with fermionic gauge charge, where the gauge charge is identified as the Bogoliubov quasi-particle\footnote{Here is another heuristic argument to see why the gauge charge should be fermionic. This argument is similar in spirit to the argument in \cite{Fidkowski2019} for the case with on-site $Z_{2}$ symmetry. We start from the $E_{8}$ root state with the chiral edge modes of the $E_{8}$ state running on the rotational axis, which we choose to be $x$-axis. Notice that, on the $x-y$ plane, the $C_{2}$ rotation becomes the reflection symmetry. Now we introduce a pair of 3-fermion topological orders on the $x-y$ plane, related by $C_{2}$ rotation. We choose the edge chirality of the 3-fermion topological order edges to be opposite to the $E_{8}$ edge. The result in Ref.~\cite{song17topological} implies that, on the surface of a 3D reflection SPT state, the chiral edge modes of the 3-fermion topological order and the chiral edge modes of the $E_{8}$ state with opposite chirality can be gapped out while preserving the reflection symmetry. Focusing on the $x-y$ plane with $C_{2}$ symmetry, we have essentially the same setting. Therefore, we obtain a state with a 3-fermion topological order on the $x-y$ plane. To produce an isotropic 3D topologically ordered state, we suppose that the 3D bulk is the $Z_{2}$ gauge theory with fermionic gauge charge. The the 3-fermion topological order on the $x-y$ plane can then be made trivial by condensing the bound state of fermionic gauge charge and one of the fermionic anyons in the 3-fermion topological order. This results in the desired $Z_{2}$ gauge theory with fermionic gauge charge in the bulk.}. This conclusion is consistent with the case of on-site $Z_{2}$ symmetry.

The anomalous nature of this $Z_{2}$ gauge theory is hidden in the structure of the loop excitation. To see this, we first notice that the same $Z_{2}$ gauge theory can be obtained from Eq.~\ref{eqn:cluston} by using  the ``vortex condensation" argument\cite{Chong2013,Metlitski2015}. We start by introducing the same pairing term into the cluston theory. Now the $U(1)$ symmetry is broken and the boundary is in the superconducting state. We would like to restore the $U(1)$ symmetry by proliferating vortices while maintaining the pairing gap. To do so, we need to understand the structure of the vortices in the superconducting state. One possible way to proceed is to solve the vortices explicitly. However, this is not necessary since, in Sec.~\ref{sec:sft}, we have shown that Eq.~\ref{eqn:cluston} can be dimensionally reduced to a pair of counter-propagating chiral modes, described by Eq.~\ref{eqn:1dchiral}, on the rotational axis, and the procedure of constructing a vortex in the superconducting state is essentially the same as a procedure of the two-step dimensional reduction. We expect that the results of two different dimensional reduction procedure are adiabatically connected. Therefore, the gapless modes in the core of vortex must be equivalent to the charge-1-charge-3 helical modes described by Eq.~\ref{eqn:1dchiral} in order to match the bulk invariant. We see that there is an obstruction to enter a symmetry-preserving gapped state by proliferating vortices due to the presence of these helical gapless modes in the cores of vortices. 

The $\zz$ classification of the $E_{8}$ root state would imply that the gapless modes in the two-fold vortex loops can be gapped out while preserving the symmetry. To see the classification is indeed $\zz$, we use the following trick. We start from the 3D boundary of the $E_{8}$ root state with the chiral edge modes of the $E_{8}$ state on the $x$-axis. Then, we compress the system in the $z$ direction to obtain a 2D system and the $C_{2}$ rotation becomes reflection symmetry. This 2D system is now essentially the same as the 2D surface of a 3D reflection SPT state with an $E_{8}$ on the mirror plane. It has been shown in Ref.~\cite{song17topological} that stacking two copies of such surfaces results in a trivially gapped state. This suggests that the classification is indeed $\zz$. We can therefore condense two-fold vortices to produce an insulating state and restore the $U(1)$ symmetry. The resulting insulating state is a $Z_{2}$ gauge theory with fermionic gauge charge tensor with a trivial fermion, where the $U(1)$ symmetry only acts on the trivial fermion. We can now gauge the $U(1)$ symmetry such that this trivial fermion is excluded from the excitation spectrum and only the $Z_{2}$ gauge theory remains. Following the above reasoning, we see that there are helical gapless modes, consisting of charge-1 and charge-3 fermions described by Eq.~\ref{eqn:1dchiral}, that are equivalent to the gapless modes of the $E_{8}$ edge state in the core of the loop excitation, which is the descendant of the fundamental vortex. This is one of the anomalous signature of this $Z_{2}$ gauge theory.

From Eq.~\ref{eqn:3dc2-1} and Eq.~\ref{eqn:3dc2-2} we see that $C_{2}$ squares to $-1$ on the gauge charge since it is identified as the Bogoliubov quasi-particle. This property of the gauge charge can also be seen from the compression argument. Again, we compress the system in the $z$ direction to obtain a 2D system, on which the $C_{2}$ rotation becomes the reflection symmetry, and we obtain a surface of a 3D reflection SPT state with an $E_{8}$ on the mirror plane. Ref.~\cite{song17topological} also shows that the surface topological order of a 3D reflection SPT state built form an $E_{8}$ state is a 3-fermion $Z_{2}$ gauge theory with $e_{f}Pm_{f}P$ symmetry fractionalization pattern. This means that the reflection symmetry $M^{2}=-1$ on gauge charge and gauge flux. When we compress our 3D $Z_{2}$ gauge theory, the resulting 2D $Z_{2}$ gauge theory is precisely the $e_{f}Pm_{f}P$ state. In particular, the gauge charge in the 3D $Z_{2}$ gauge theory just becomes the gauge charge in the $e_{f}Pm_{f}P$ state. Therefore, the gauge charge in our 3D $Z_{2}$ gauge theory must carry half $C_{2}$ charge. 

The fact that the gauge charge carries half $C_{2}$ charge has a dramatic consequence on the $C_{2}$ defect loop $\Omega$. The defining property of a $C_{2}$ defect loop is that, when a gauge charge braids with the defect loop, it implements an $C_{2}$ transformation on the gauge charge. Since $C_{2}$ squares to $-1$ on the gauge charge, we see that the defect loop $\Omega$ must satisfy the fusion rule: $\Omega \times \Omega = m$, where $m$ represents the loop excitation in the gauge theory. Since we have shown that the loop excitation $m$ carries gapless modes that are equivalent to the $E_{8}$ edges. The $C_{2}$ defect loop $\Omega$ should carry gapless modes that are equivalent to $c=4$ $SO(8)_{1}$ chiral CFT. This is another anomalous signature of this $Z_{2}$ gauge theory.

\section{Discussion}
In this paper, we classified 4D bosonic $C_{2}$ SPT phases by using the topological crystal approach. The classification is found to be $\zz \times \zz$. This classification is consistent with the case of on-site $Z_{2}$  symmetry as expected from the crystalline equivalence principal. One of the $\zz$ root state is understood as having a Ising SPT state on the 2D $C_{2}$ invariant plane. The other $\zz$ root state is given by having a $E_{8}$ state on the rotational invariant plane. This state is beyond-cohomology since the building block itself is not an SPT state classified by group cohomology. 

Focusing on this $E_{8}$ root state, we consider its boundary field theories. We found a variant of QED4 with a single charge-1 and a single charge-3 Dirac fermions with the $C_{2}$ symmetry action defined in Eq.~\ref{eqn:3dc2-1}-\ref{eqn:3dc2-2} is a candidate boundary field theory by using the dimensional reduction argument. This field theory is inspired by its $(2+1)$d version based on the ``cluston" construction, introduced in Ref.~\cite{Chong2015cluston} for various SPT phases with time-reversal symmetry in 3 spatial dimension. We show that its $(3+1)$d generalization can describe the boundary of 4D bosonic $C_{2}$ SPT phases built by placing an $E_{8}$ state on the rotational invariant plane. 

We further consider how to obtain a gapped topologically ordered state from the field theory. The topological order we obtain is a $3D$ $Z_{2}$ gauge theory with fermionic gauge charge. One of the anomaly signature of this $Z_{2}$ gauge theory is shown in the core of the loop excitation--- there are gapless modes carry the same anomaly as the edge modes of the $E_{8}$ state in the core of the loop excitation. Another anomaly signature is that the $C_{2}$ defect loop carries gapless modes that are equivalent to $c=4$ $SO(8)_{1}$ chiral CFT.

The QED4 with a charge-1 and a charge-3 Dirac fermions might also describe the beyond-cohomology state with on-site $Z_{2}$ symmetry. It would be interesting to show this explicitly by using more traditional field theory analysis. In general, it would be interesting to have a more detailed understanding of 4D SPT phases with on-site and/or crystalline symmetry. Although the 4D SPT phases themselves are unrealistic, it's crucial for understanding the anomalies of $(3+1)$d field theories. 

It's straightforward to see that the same $E_{8}$ root state exist if the symmetry is $C_{n}$ rotation since there is still a 2D  $C_{n}$ invariant plane which can support the $E_{8}$ state. Indeed, if we consider the case with on-site $Z_{2}$ symmetry, the beyond-cohomology state is the descendant of the one with $U(1)$ symmetry. The similar beyond-cohomology phases exist if the symmetry is broken down to its $Z_{n}$ subgroup. The classification was found to be $\z$ for $U(1)$ symmetry and is $\z_{n}$ for $Z_{n}$ symmetry in Ref.~\cite{Wen2015}. We expect the same classification for the SPT phases with $C_{n}$ rotation based on the crystalline equivalence principal. The boundary theories are presumably similar as we discussed in this paper but we leave the detailed consideration to the future work. 

\begin{acknowledgments}
I'm grateful to Meng Cheng for illuminating discussions on the property of the $C_{2}$ defect loop in the anomalous $Z_{2}$ gauge theory, to Chong Wang for discussions on the gapless modes in the vortex core, to Maissam Barkeshli, Lukasz Fidkowski, and Cenke Xu for useful correspondence. S.-J.H. acknowledges support from a JQI postdoctoral fellowship and the Laboratory for Physical Sciences. 
\end{acknowledgments}

\bibliographystyle{apsrev4-1}
\bibliography{4dc2spt}

\end{document}